\documentclass[aps,prl,reprint,superscriptaddress]{revtex4-1}

\usepackage{graphicx}
\usepackage{dcolumn}
\usepackage{bm}

\usepackage{amsmath}

\graphicspath{{./figures/}}

\usepackage[usenames, dvipsnames]{color}

\DeclareMathOperator{\ud}{d\!}

\begin{document}

\preprint{APS/123-QED}

\title{Dynamic self-assembly of microscale rotors and swimmers}

\author{Megan S.\ Davies Wykes}
\email{megan.davieswykes@cantab.net}
\affiliation{Applied Mathematics Laboratory, Courant Institute, New York University}%

\author{J\'{e}r\'{e}mie Palacci}%
\email{palacci@physics.ucsd.edu}
\affiliation{Department of Physics, UC San Diego}%
\affiliation{Applied Mathematics Laboratory, Courant Institute, New York University}%

\author{Takuji Adachi}
\affiliation{Molecular Design Institute, Department of Chemistry, New York University}

\author{Leif Ristroph}
\affiliation{Applied Mathematics Laboratory, Courant Institute, New York University}

\author{Xiao Zhong}
\affiliation{Molecular Design Institute, Department of Chemistry, New York University}

\author{Michael D.\ Ward}
\affiliation{Molecular Design Institute, Department of Chemistry, New York University}

\author{Jun Zhang}
\affiliation{Applied Mathematics Laboratory, Courant Institute, New York University}
\affiliation{Department of Physics, New York University}
\affiliation{NYU-ECNU Institutes of Mathematical Sciences and Physics Research, NYU-Shanghai}

\author{Michael J.\ Shelley}
\affiliation{Applied Mathematics Laboratory, Courant Institute, New York University}

\date{\today}

\begin{abstract}
	
Biological systems often involve the self-assembly of basic components into complex and functioning structures. Artificial systems that mimic such processes can provide a well-controlled setting to explore the principles involved and also synthesize useful micromachines. Our experiments show that immotile, but active, components self-assemble into two types of structure that exhibit the fundamental forms of motility: translation and rotation. Specifically, micron-scale metallic rods are designed to induce extensile surface flows in the presence of a chemical fuel; these rods interact with each other and pair up to form either a swimmer or a rotor. Such pairs can transition reversibly between these two configurations, leading to kinetics reminiscent of bacterial run-and-tumble motion.

\end{abstract}

\maketitle



Self-assembly is a hallmark of biological systems as organisms construct functional complex materials and structures from simpler components. The use of self-assembly in the fabrication of new materials is attractive owing to its inherent versatility and potential for mass production \cite{Whitesides2002}. Research has focused primarily on systems where the formation of macrostructures does not require the continuous input of energy \cite{Zhang2014,Sacanna2010,Chen2011a,Wang2012}, a process known as equilibrium self-assembly. Another route to self-assembly is dynamic: where structures persist only while energy is being supplied to the system \cite{Wang2015}. Artificial micron-scale motors, immersed in a fuel laden fluid, have been shown to spontaneously self-organize into crystal structures \cite{Palacci2013}, and form into asters of two or more particles \cite{Ahmed2014}. These studies have active components, producing a local fluid flow which leads to motility. A striking aspect of these active systems is their potential for emergent dynamics, whereby individuals and groups have qualitatively different behaviour \cite{Saintillan2013}, such as flocking \cite{Bricard2013}, formation of flow structures on larger scales than the individual components \cite{Sanchez2012}, or collective flows which are faster than those induced by individual active particles \cite{Saintillan2012}. 

\begin{figure}[tb]
	\centering
	\small
	\begin{picture}(230,120)
	\put(0,0){\includegraphics[height=55px]{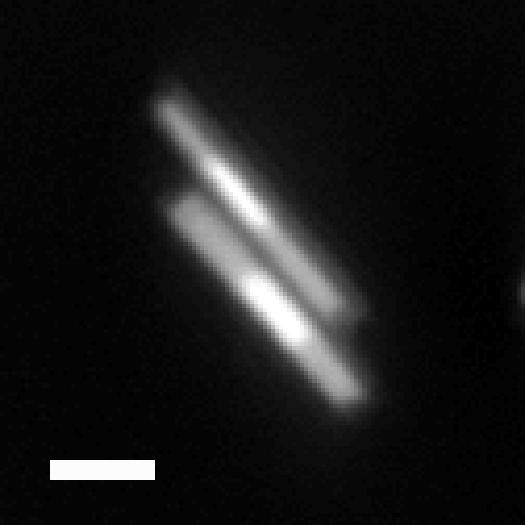}}
	\put(40,5){\textcolor{White}{(d)}}
	\put(58,0){\includegraphics[height=55px]{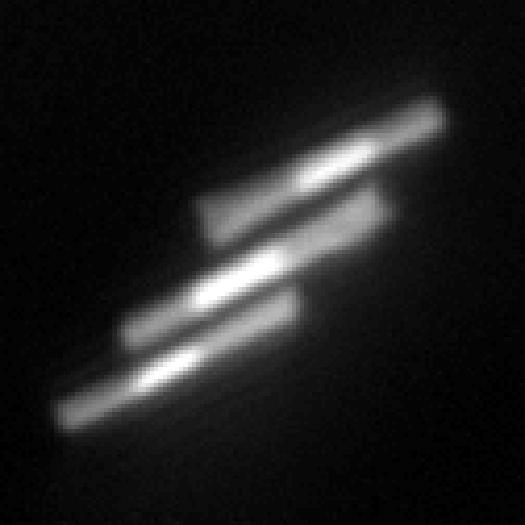}}
	\put(98,5){\textcolor{White}{(e)}}
	\put(116,0){\includegraphics[height=55px]{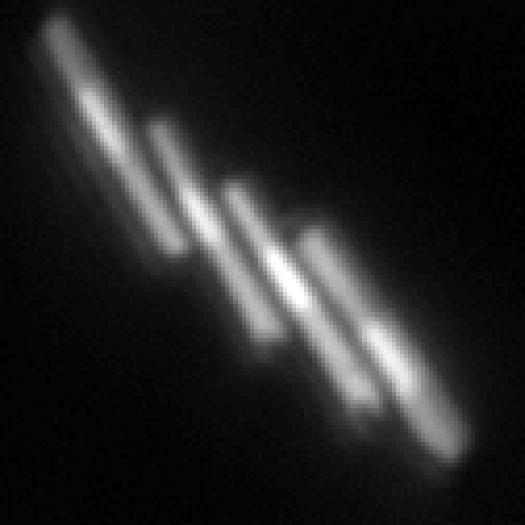}}
	\put(120,5){\textcolor{White}{(f)}}
	\put(174,0){\includegraphics[height=55px,angle=90]{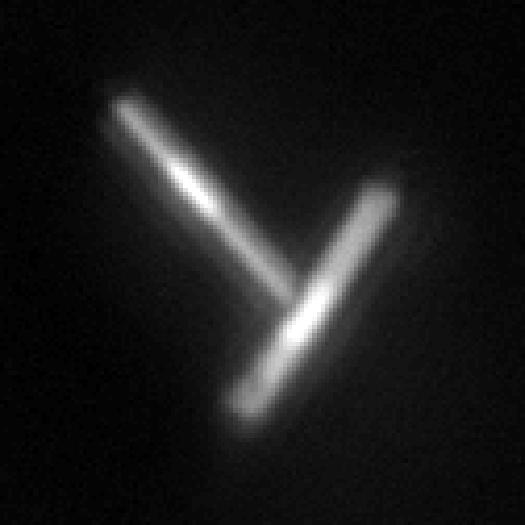}}
	\put(215,5){\textcolor{White}{(g)}}
	\put(0,60){\includegraphics[height=55px]{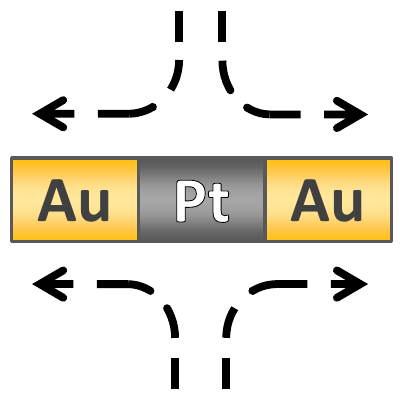}}
	\put(58,60){\includegraphics[height=55px]{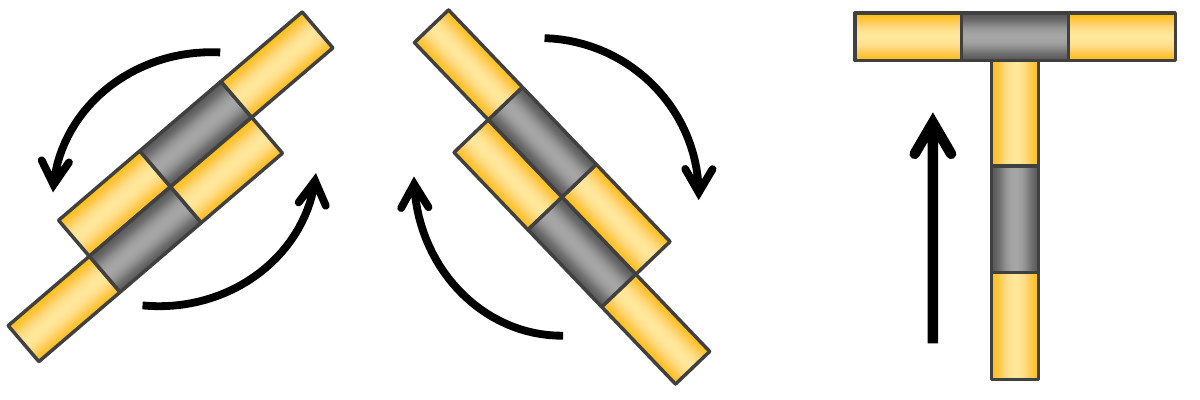}}
	\put(5,64){(a)}
	\put(103,65){(b)}
	\put(65,110){R1}
	\put(147,110){R2}
	\put(215,65){(c)}
	\end{picture}
	
	\normalsize
	\caption{\label{fig:diagram_and_micrographs} (a) Schematic of tripartite rod and expected extensional flow. Motion of (b) rotors in arrangement R1 and R2, (c) T-swimmers (vertical rod is the leg, horizontal rod is the top). Micrographs of (d-f) rotors, (g) T-swimmer. Scale bar $1 \mu$m refers to all micrographs.}
\end{figure}
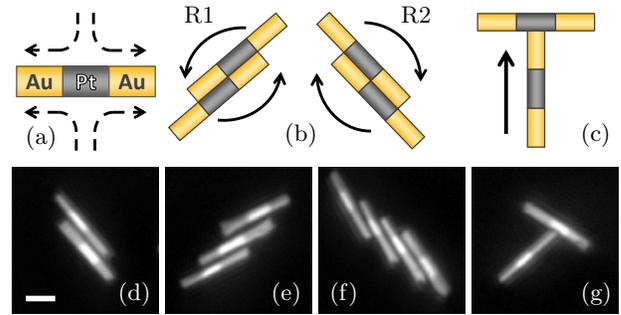

Here we describe how two of the simplest machines can be formed by dynamic self-assembly from a single type of building block. Active, but immotile, rods spontaneously self-assemble into structures that exhibit the two fundamental types of motion: rotation and translation (Fig.\ \ref{fig:diagram_and_micrographs}d-g). This occurs without imposing any external field. Assemblies can transition between rotor and swimmer, leading to behaviour reminiscent of the run-and-tumble motion of \emph{E. coli} \cite{Berg2005}. The type and direction of motion is determined by the configuration of rods in an assembly, thereby linking the self-assembly to the emergent motility. This is a rare example of artificial dynamic self-assembly and therefore represents an important step towards making more complex micron-scale machines.

\section{System and phenomenology}

The particles described herein are constructed of three metal sections, in the arrangement gold-platinum-gold (Au-Pt-Au). These micron-scale rods are expected to produce an extensile flow when submerged in a solution of chemical fuel (Fig.\ \ref{fig:diagram_and_micrographs}a). The particles are dense compared to the solution, therefore in experiments they sit close to the lower surface and are essentially confined to 2D. These particles are `active' in that they generate a flow, but are immotile as this flow is symmetric and asymmetric motion is required to swim at low Reynolds number. Although many studies have examined self-assembly for motile particles, there have been few experimental studies on the self-assembly of active, but immotile particles. Experiments of silver nanorods surrounded by contractile flows (the reverse of those studied here), induced by an external electric field that also aligns the nanorods, found evidence of hydrodynamic pair interactions \cite{Rose2009}. Theoretical studies of `extensors' have indicated the presence of interesting collective behaviour \cite{Saintillan2007,Pandey2014}. In the case of individually immotile particles, motility itself can be a signature of emergent dynamics as it indicates some form of symmetry breaking in the system.

The system described here builds on research into micron-scale bipartite gold-platinum (Au-Pt) rods, which swim by self-electrophoresis when placed in a solution of hydrogen peroxide (H$_2$O$_2$) fuel \cite{Paxton2004,Wang2006,Takagi2014,Wang2009,Ebbens2010,Takagi2013,Moran2010}. Electro-chemical decomposition of H$_2$O$_2$ results in a gradient in proton concentration, which corresponds to an electric field pointing from Pt to Au \cite{Wang2013a}. The rods themselves have an overall negative charge. Therefore, the positively charged electrical double layer surrounding the rod experiences a force due to the self-generated electrical field \cite{Moran2010,Moran2011,Wang2013a}. A fluid flow develops on the rod surface, from Pt to Au, causing the rod to swim with its Pt end leading. Swimming rods have been observed to form pairs transiently \cite{Wang2013}, suggesting the presence of attractive interactions. The composition of particles can be tailored, thereby altering the flow configuration around a rod.



\begin{figure}[tb]
	\includegraphics[width=0.48\textwidth]{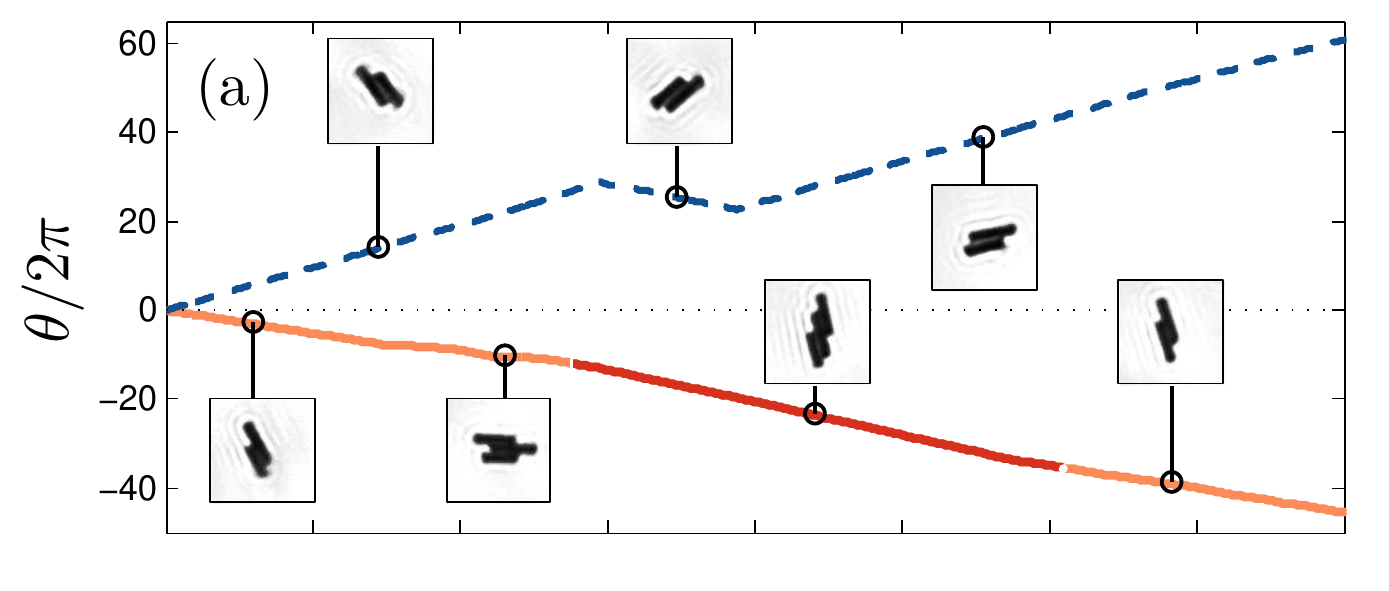}%
	\vspace*{-5.556mm}
	
	\includegraphics[width=0.48\textwidth]{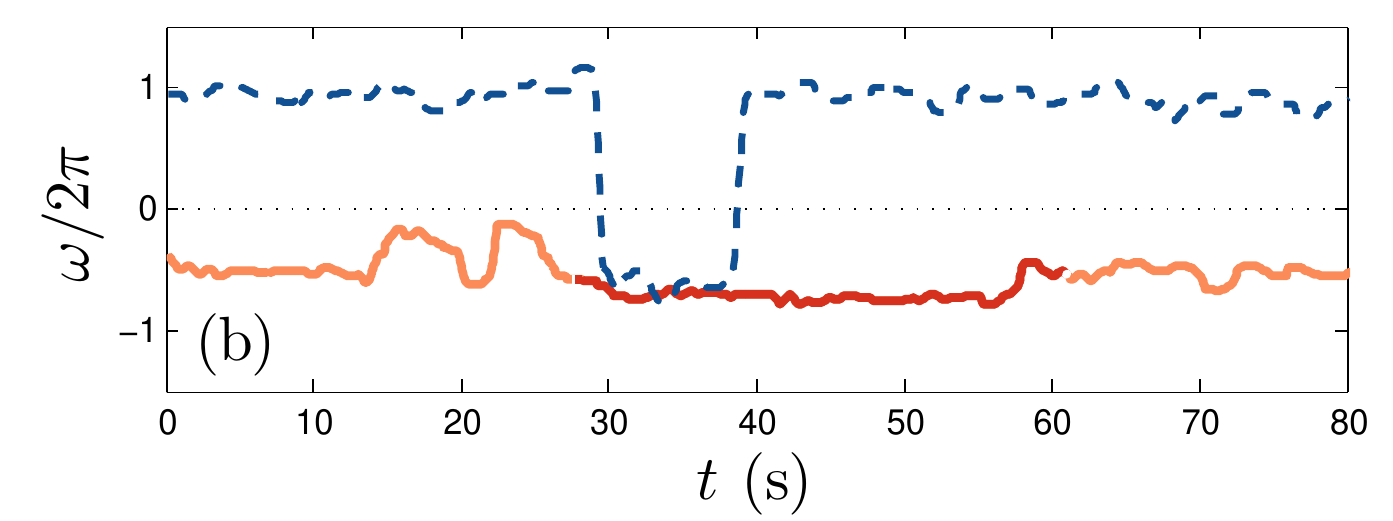}%
	
	\caption{\label{fig:change_direction} Rotor which changes direction (blue, dashed) and a rotor that has a rod added to it (red, orange): the rotor begins with two rods (orange), separates and re-forms with three rods (red). The three-rod assembly is initially in a staggered arrangement (red, dotted), then adjusts to become a three-rod rotor (red, solid), then loses a rod, becoming a two-rod rotor (orange). Plotted are (a) the total number of rotations $\theta/2\pi$ (winding number), (b) the rotation rate, $\omega/2\pi$ (Hz). Counter-clockwise is positive.}
\end{figure}

Tripartite rods of length $2.18 \pm 0.4 \mu$m and diameter $0.34 \pm 0.08 \mu$m are fabricated as described in the Supplementary Information (SI). The motion of these particles on a glass coverslip is observed using an inverted optical microscope, with a $100\times$ objective (oil immersion, N.A.\ 1.4) and diascopic illumination. Reflected illumination can be used to identify the individual metal segments as visible on Fig.\ \ref{fig:diagram_and_micrographs}d-g. When these tripartite rods are placed in a solution of H$_2$O$_2$ they spontaneously self-assemble into several distinct configurations.

\begin{figure}[tb]
	
	\centering
	\begin{picture}(180,140)
	\put(30,0){\includegraphics[height=95px]{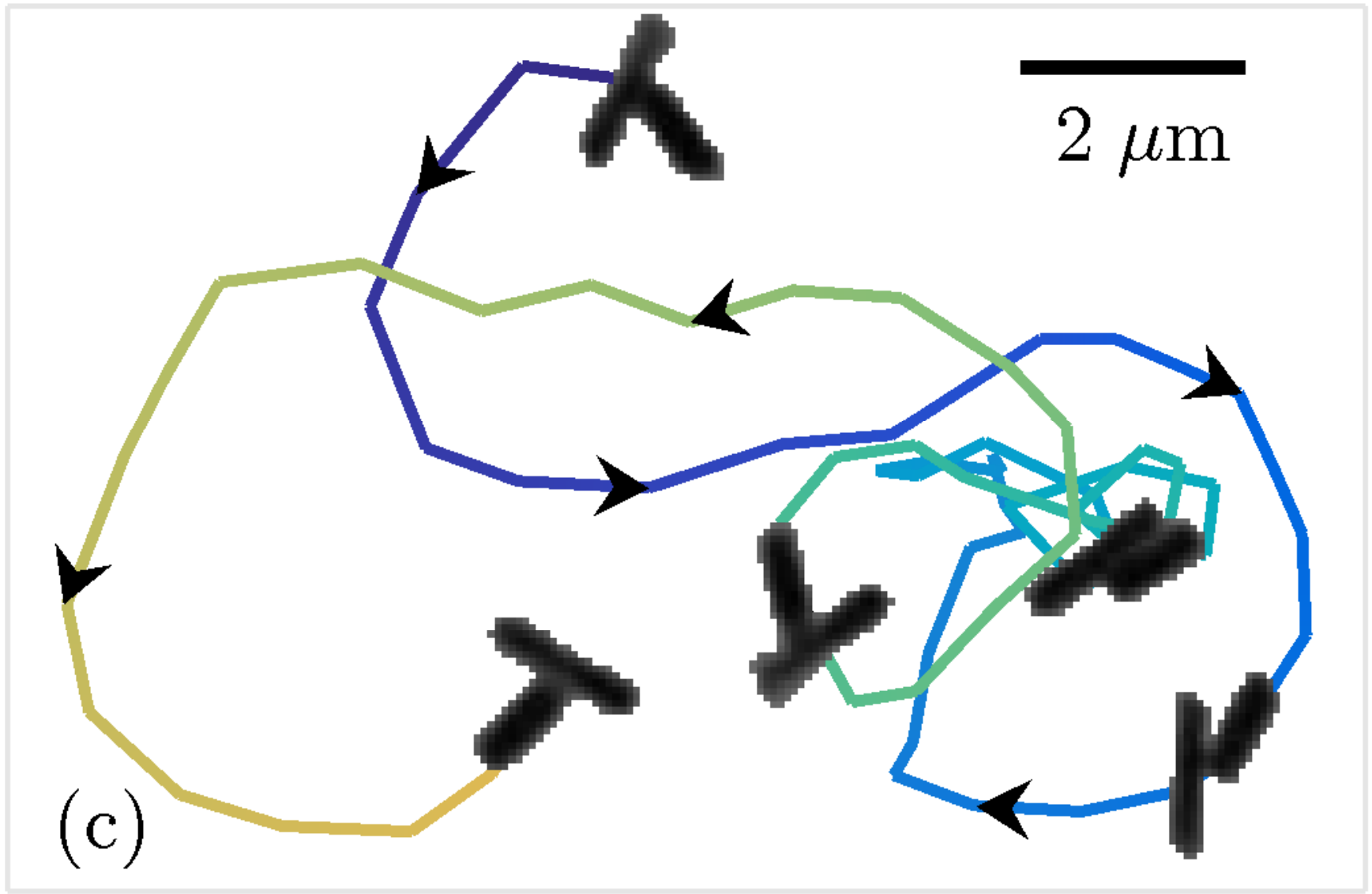}}
	\put(0,70){\includegraphics[height=65px]{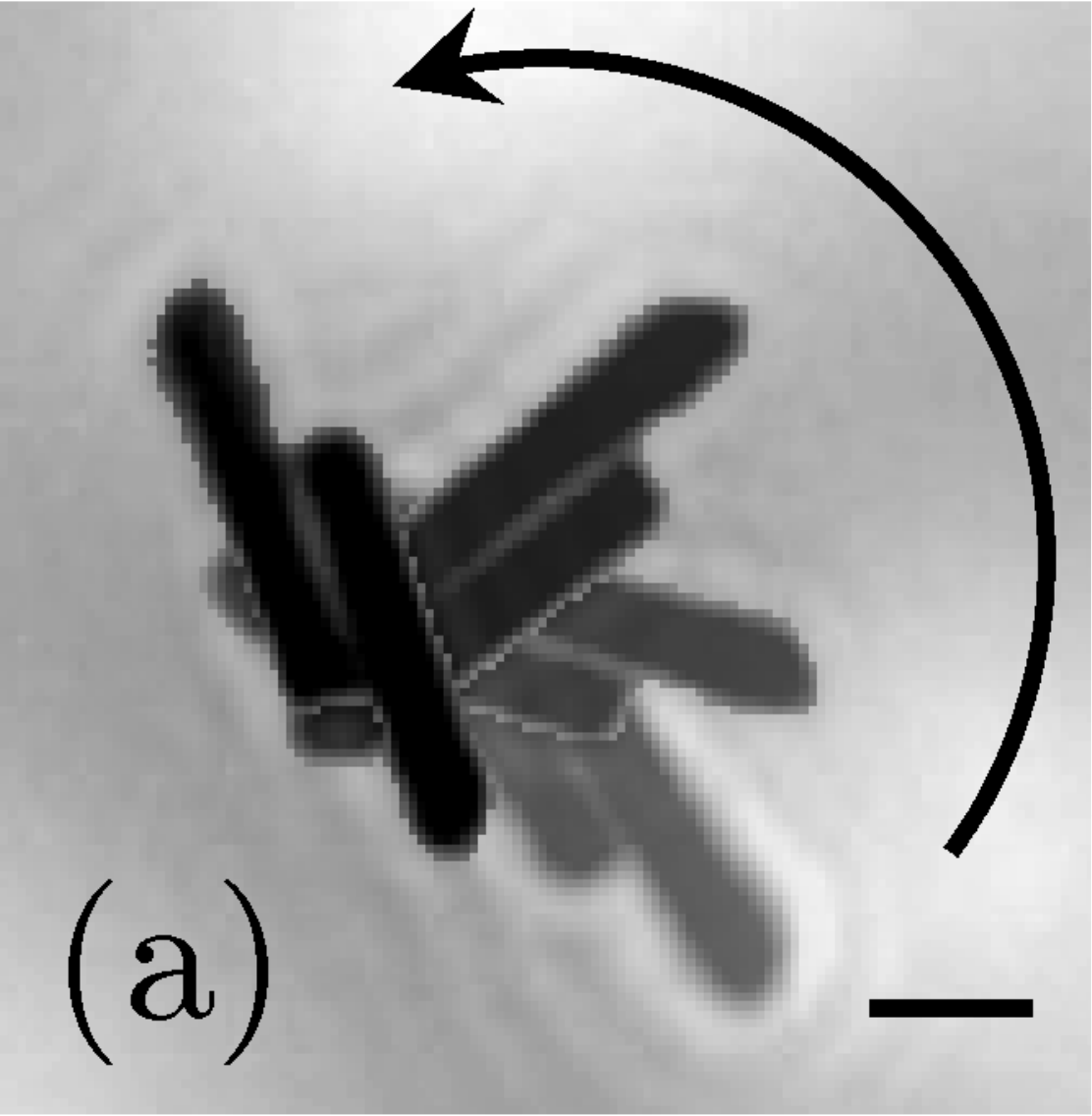}}
	\put(67,98){\includegraphics[height=37px]{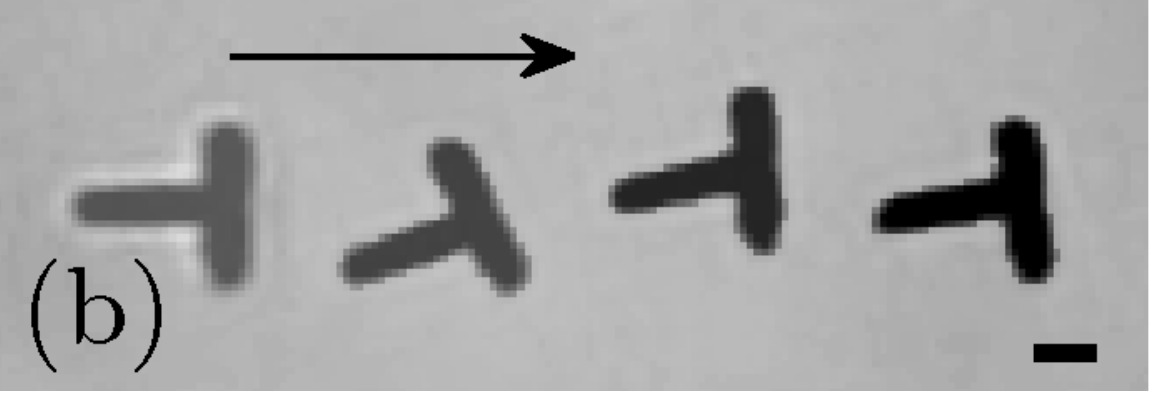}}
	\end{picture}
	
	\caption{\label{fig:motion}  Overlay images of (a) rotor, time interval: 0.5 s, (b) T-swimmer, time interval: 1 s. Scale bars $1\mu$m. (c) Path taken by a pair of rods transitioning from \mbox{T-swimmer}, to rotor, to T-swimmer (scale bar refers to path, overlay images at $\frac{1}{2}$ scale). Overlay images at $t = 0$, $1.6$, $4.4$, $5$, and $7.2$ seconds.}
\end{figure}

The attraction between rods appears short-ranged, on the order of a rod length, as further separated rods appear to act diffusively. The most common assembly consists of two or more parallel and regularly staggered rods, as seen in Fig.\ \ref{fig:diagram_and_micrographs}d-f. An example of rotor self-assembly is included as supplementary movie S1. Once assembled, these structures rotate at a steady rate and in a direction determined by the rod configuration (Fig.\ \ref{fig:motion}a). As the rods are dense compared to the solution, they are essentially confined to 2D and consequently rotors are chiral. There are two possible arrangements for the particles in a rotor, these will be referred to as R1 and R2, as illustrated by Fig.\ \ref{fig:diagram_and_micrographs}b. Rotors always rotate with the outer rod leading, i.e.\ R1 rotates counter-clockwise and R2 rotates clockwise. Individual rotors have mean rotation rates in the range $\omega = 0.4 - 2$Hz for $3\%$ H$_2$O$_2$ (SI). While some rotors breakup after short times ($\sim 10$s), others persist without breakup for longer than the recording time of experiments ($> 120$s) and are sufficiently stable to survive collisions with other rods. 

\begin{figure*}[hbt]
	\includegraphics[height=0.2\textwidth,clip]{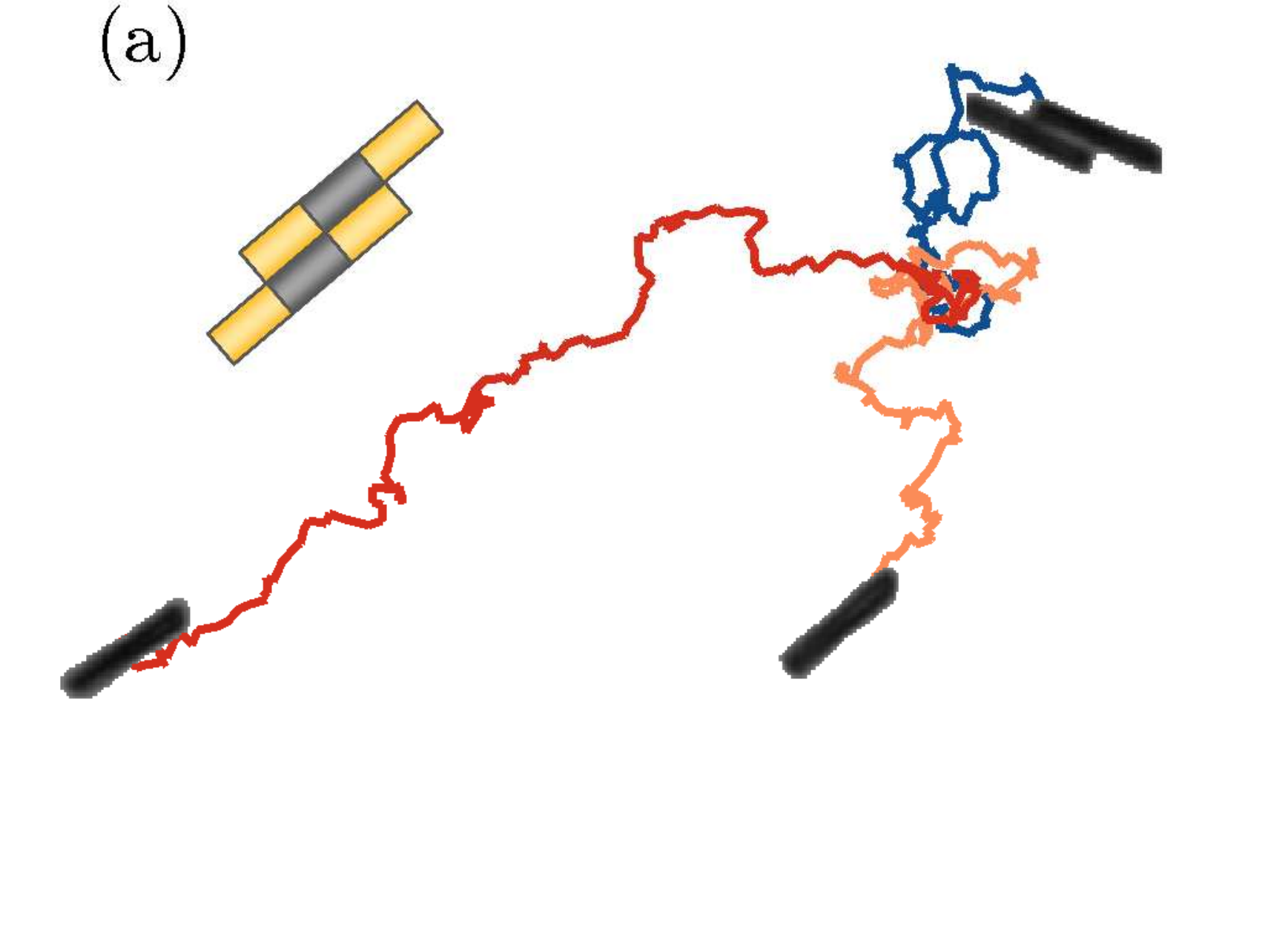}	
	\includegraphics[height=0.2\textwidth,clip]{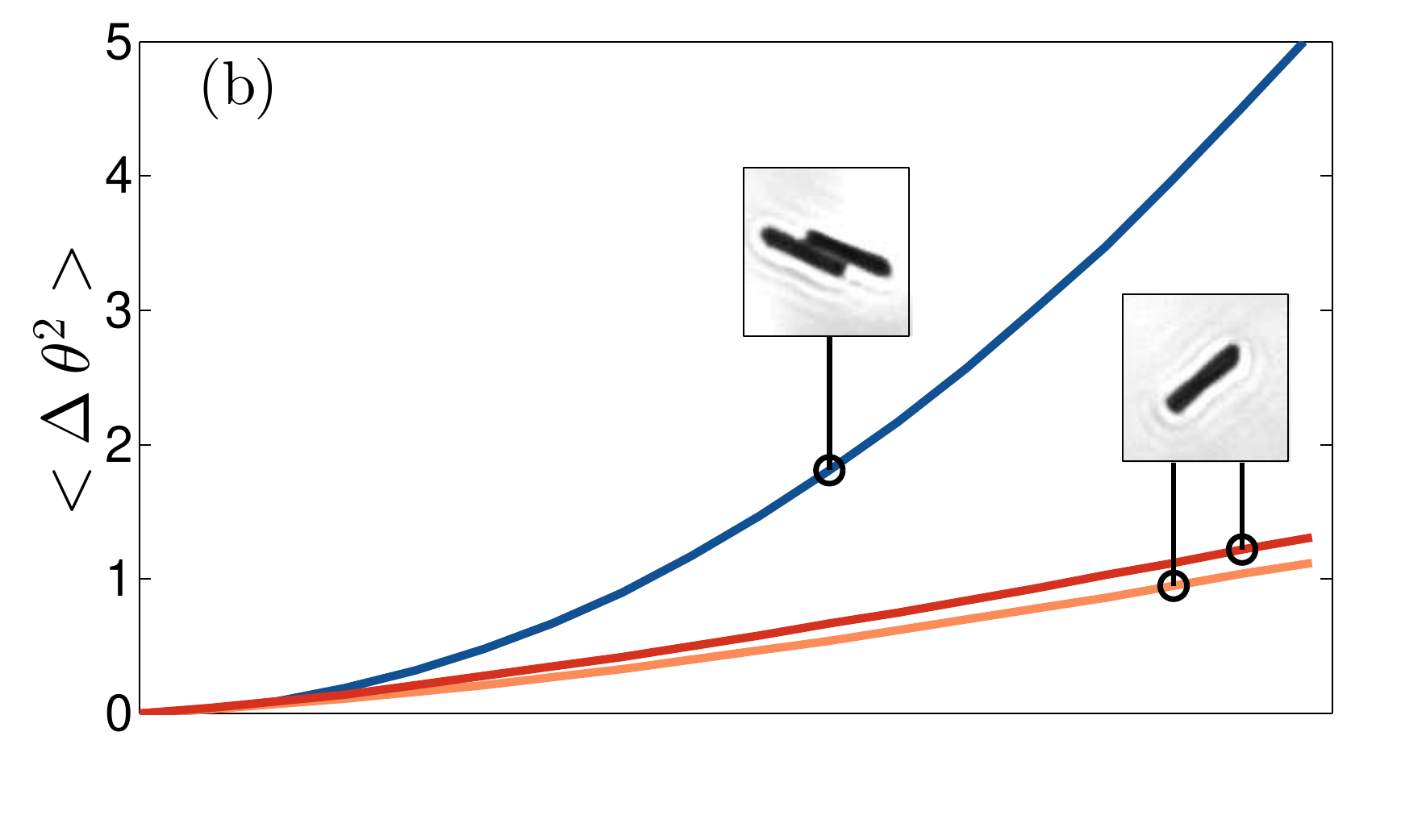}%
	\includegraphics[height=0.2\textwidth,clip]{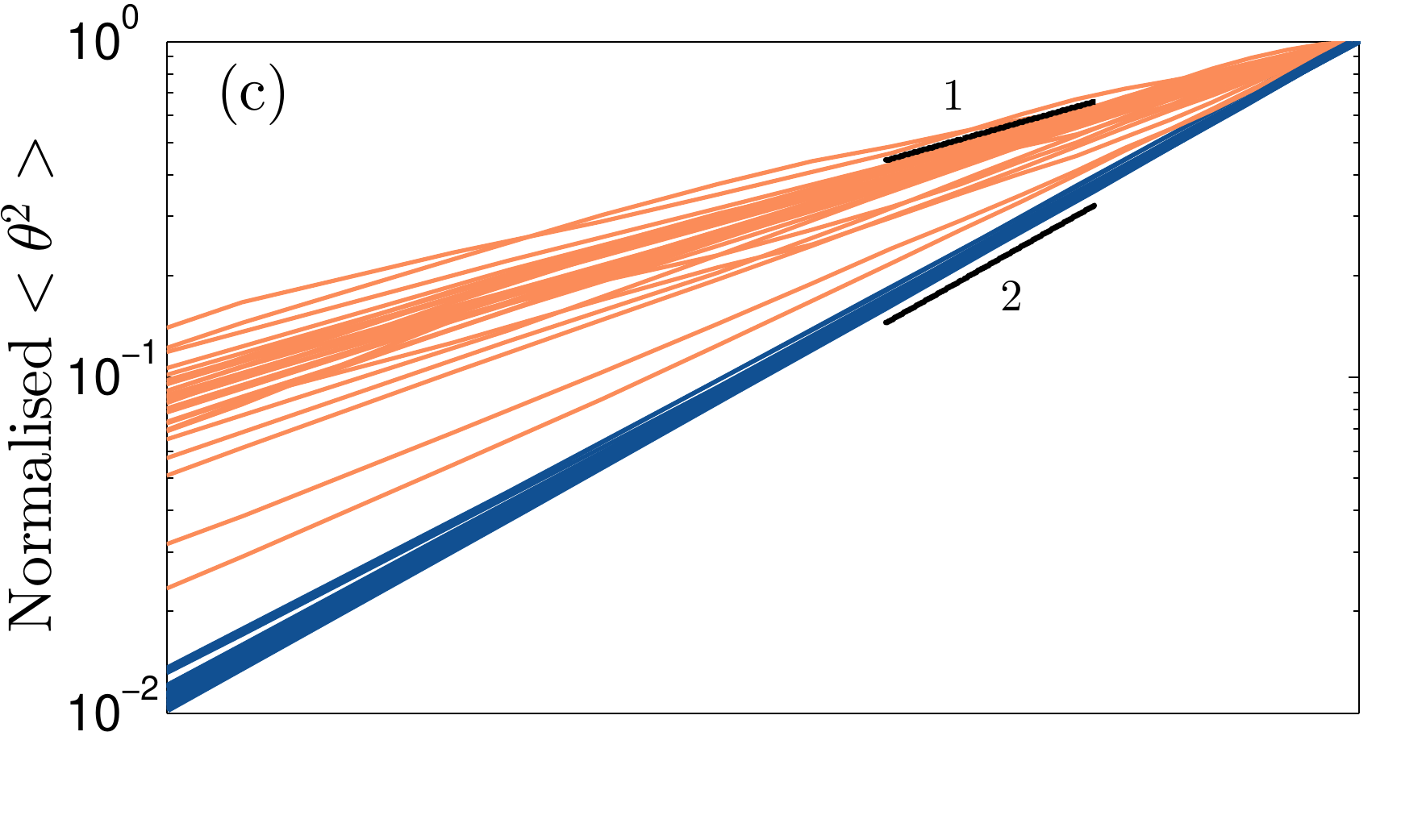}%
	
	\vspace*{-7.5mm}
	
	\includegraphics[height=0.2\textwidth,clip]{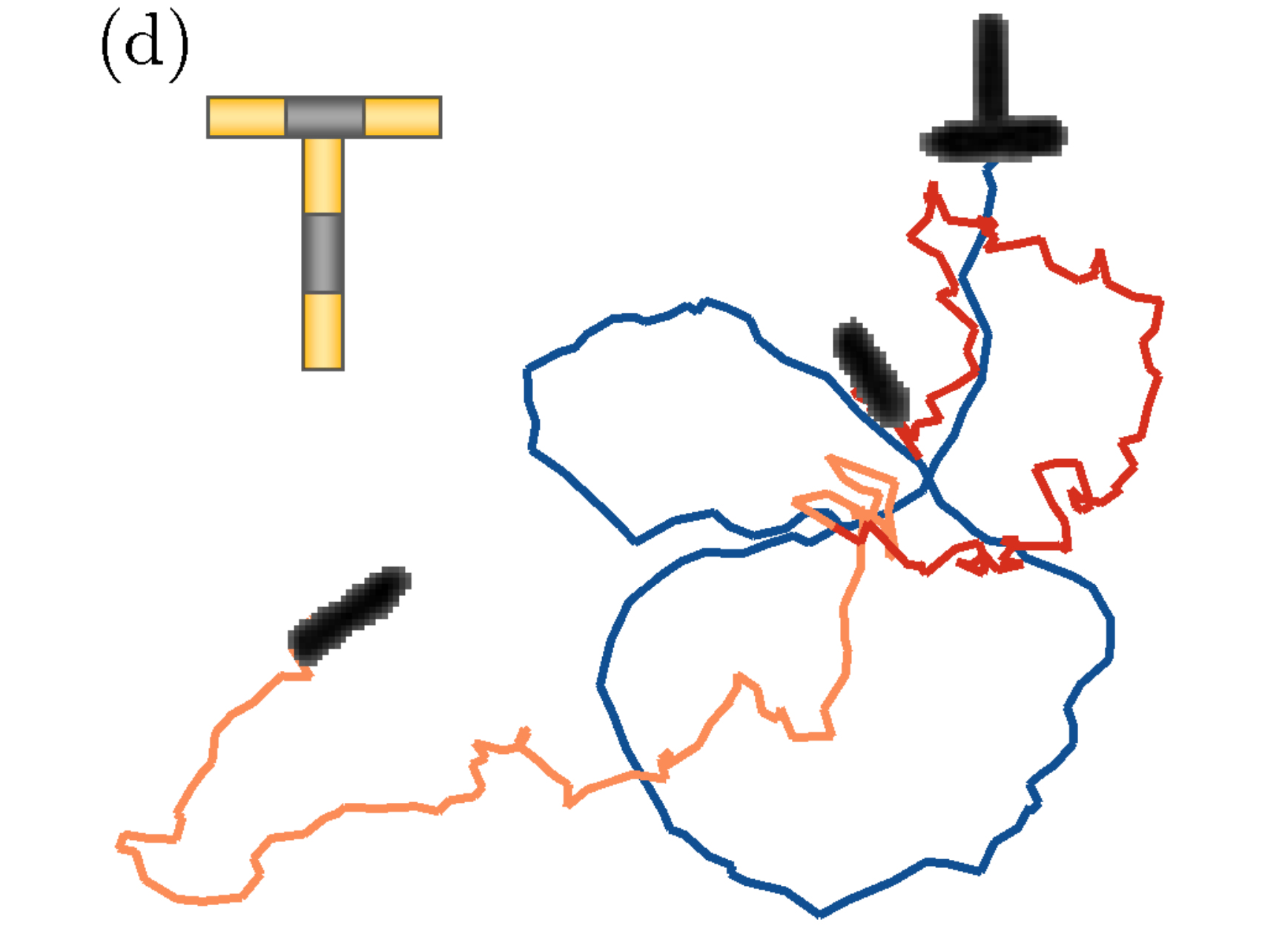}
	\includegraphics[height=0.2\textwidth,clip]{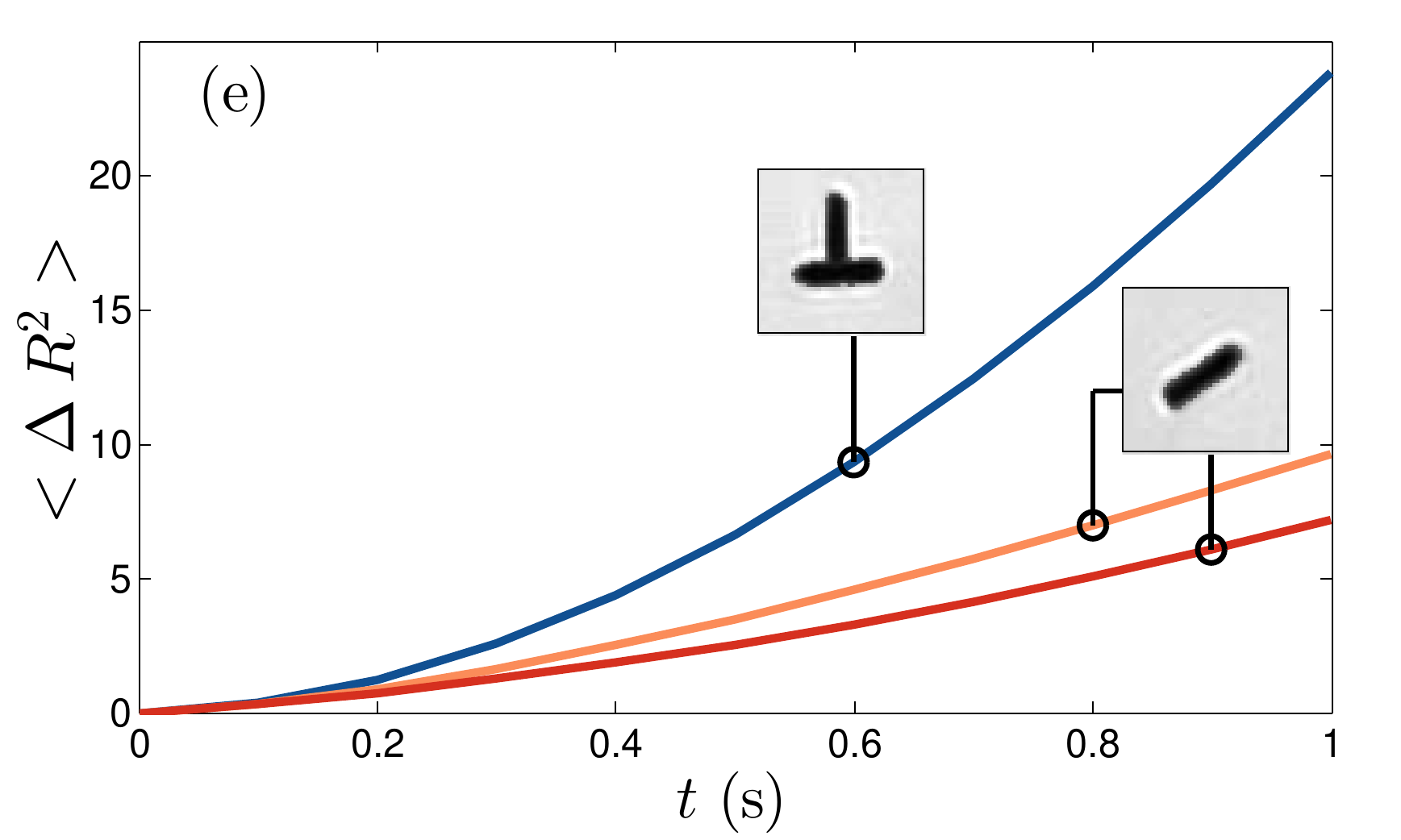}%
	\includegraphics[height=0.2\textwidth,clip]{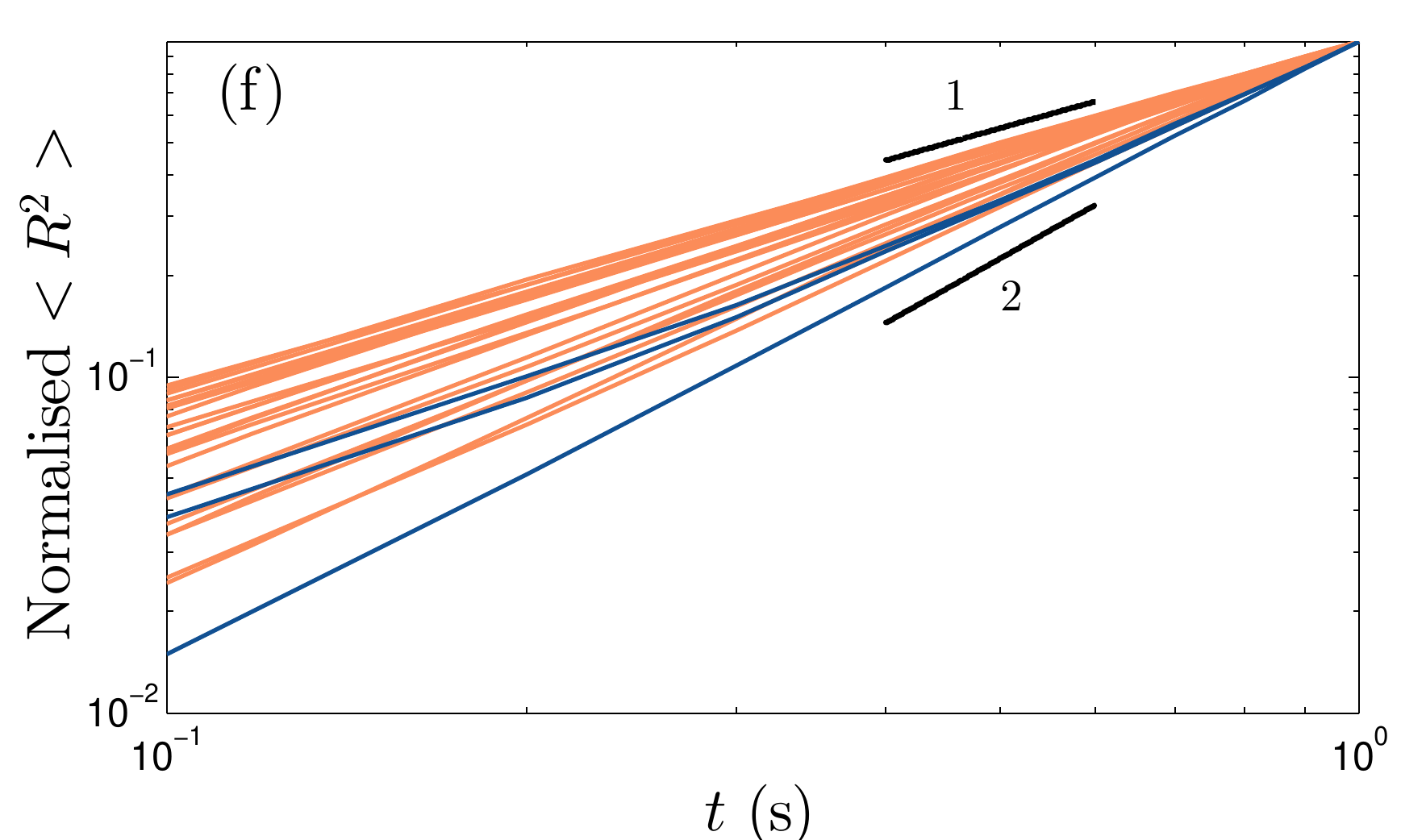}%
	
	\caption{\label{fig:emergent_dynamics} Emergent dynamics of a rotor and a T-swimmer: (a) Tracks of rotor (blue) and component rods after breakup (orange, red) plotted for 10s. (b) MSAD (rad$^2/$s$^2$) for rotors and rods from (a). (c) MSAD (normalised by maximum value) for 40 rotors (blue) and 29 single rods (orange). (d) Tracks of T-swimmer (blue) and component rods after breakup (orange, red) plotted for 10s. (e) MSD ($\mu$m$^2/$s$^2$) for T-swimmer and rods from (d). (f) MSD (normalised by maximum value) against time for 3 T-swimmers (blue) and 33 single rods (orange). The majority of single rods have slightly super-diffusive translation and rotation, consistent with slight motility. Rotor MSAD shows a quadratic dependence on time, characteristic of powered rotation. T-swimmers show a mix of translational diffusion and powered translation. Overlaid micrographs on (a,d) are to scale with particle tracks.}
\end{figure*}

Assembly is reversible -- pairs can separate and re-form -- and the reassembled structures rotate at an identical speed and in the same direction, if assembled in the original configuration. In Fig.\ \ref{fig:change_direction}a (dashed line) an example is shown where a rotor begins in configuration R1, rotating counter-clockwise (defined as positive). At 30s, thermal fluctuations cause the rods within a rotor to break alignment. When the rotor reforms, it has switched to configuration R2 and starts rotating in a clockwise direction. At 40s, the rotor switches direction for a second time, returning to the original R1 configuration and rotating counter-clockwise. The rotation speed of this example is plotted in Fig.\ \ref{fig:change_direction}b (dashed line) and is equal to 1 Hz when rotating counter-clockwise and 0.5 Hz when rotating clockwise. However, many rotors rotate at an identical speed even after a change of direction. An example of a rotor changing direction is included as movie S2. When a rod is added to a rotor this results in a small change in rotation rate. An example of this is shown in Fig.\ \ref{fig:change_direction}, where a rotor initially composed of two rods (orange) has a rod added to it (red). Initially the three-rod assembly is in a staggered arrangement (red, dotted), but then adjusts to become a three-rod rotor (red, solid). At a later time, the assembly separates back to a two-rod rotor (orange). An example of a rod being added to a rotor is included as movie S3. The staggered arrangement (referred to as an `enterprise' in group meetings) that appears as an intermediary state is less common than the staircase arrangement seen in rotors.


Another self-assembled structure is the T-swimmer, consisting of two perpendicular rods joined by a Pt-Au junction, which swims in a direction aligned with the leg of the `T', towards the end with the junction (Fig.\ \ref{fig:diagram_and_micrographs}c and Fig.\ \ref{fig:motion}b). An example can be seen in movie S4. This shape is much less common than the rotors, accounting for only $\sim 5 \%$ of assemblies. T-swimmers persist for shorter times than rotors ($\sim 10$s) and are often a transitional state when a rotor reverses direction or otherwise breaks alignment (as in movie S2). Interestingly, pairs can transition repeatedly between the T-swimmer and rotor configurations. As a T-swimmer, the pair swims in looping trajectories, as in Fig.\ \ref{fig:emergent_dynamics}d, while a rotor essentially spins in place. Figure \ref{fig:motion}c shows the trajectory of a particle pair that starts as a T-swimmer, transitions into a rotor, and then returns to the T-swimmer configuration (see movie S5). Switching between the two states is a result of thermal fluctuations. This means that the time spent either as a rotor or a swimmer is randomly distributed. Rotation for a random length of time will cause the swimming direction to be randomised when the pair switches back to a T-swimmer. These characteristics make the trajectory of a particle pair that is switching states reminiscent of bacterial run-and-tumble behaviour.


\section{Analysing the emergent dynamics}

While we have stated that individual rods appear to act diffusively and assemblies show motility, we have as yet shown no direct evidence that this is the case. There is an established method for identifying translational motion as deterministic rather than diffusive, which examines the mean squared displacement at early times \cite{Howse2007}. Similarly, deterministic rotation can be separated from diffusive motion by examining the mean squared angular displacement \cite{VanTeeffelen2008}. For clarity, we briefly outline the classical model for the path $(x,y)$ and angle $\theta$ of particles, which are diffusing isotropically, rotating and swimming, as derived elsewhere \cite{VanTeeffelen2008}. The model is composed of the set of overdamped Langevin equations 
\begin{align}
	\frac{\ud x(t)}{\ud t} & = v \cos\left(\theta(t)\right) + \sqrt{2D}\xi_x(t), \label{eq:Langevin_x} \\
	\frac{\ud y(t)}{\ud t} & = v \sin\left(\theta(t)\right) + \sqrt{2D}\xi_y(t), \label{eq:Langevin_y} \\
	\frac{\ud \theta(t)}{\ud t} & = \omega + \sqrt{2D_\theta}\eta(t). \label{eq:Langevin_theta}
\end{align}
Here $v$ is the speed of the particle, $\omega$ is the rotation rate in rad/s, $D$ is the translational diffusion coefficient and $D_\theta$ is the rotational diffusion coefficient. The noise terms $\xi_x$, $\xi_y$ and $\eta$ are uncorrelated Gaussian processes with zero mean and unit variance \cite{VanTeeffelen2008,Marine2013}.

Equation (\ref{eq:Langevin_theta}) is decoupled from Eqs.\ (\ref{eq:Langevin_x}) and (\ref{eq:Langevin_y}) and can be integrated to give the mean-squared angular displacement (MSAD)
\begin{equation}
	\left<\Delta \theta (t)^2\right> = \omega^2 t^2 + 2D_r t \label{eq:MSAD}
\end{equation}
where $\Delta \theta (t)^2 = (\theta(t_0+t) - \theta(t_0))^2$ and $\left<\bullet\right>$ indicates an average over a time series. Equation (\ref{eq:MSAD}) shows that deterministic rotation results in a quadratic MSAD, whereas rotational diffusion results in a linear MSAD \cite{VanTeeffelen2008,Ebbens2010a,Marine2013}.

The tracks of a rotor (blue) and the component rods after breakup (red, orange) are each plotted for 10s in Fig.\ \ref{fig:emergent_dynamics}a. The MSAD reveals the fundamentally different dynamics of a pair compared to a single rod (Fig.\ \ref{fig:emergent_dynamics}b). The rotation of individual rods is diffusive, but when assembled into a rotor the MSAD is quadratic with time, characteristic of deterministic rotation. The MSAD for several rotors and single rods are plotted in Fig.\ \ref{fig:emergent_dynamics}c. The motion of rotors is dominated by deterministic rotation, while the motion of the majority of single rods is diffusive.

Solving Eqs.\ (\ref{eq:Langevin_x}-\ref{eq:Langevin_theta}) to find the mean-squared displacement (MSD) reveals the existence of several regimes \cite{Ebbens2010a}. For $t \ll 1/D_r$ and $t \ll 1/\omega$ , the MSD takes the limiting form of
\begin{equation}
\left<\Delta R (t)^2\right> = 4Dt+v^2t^2
\end{equation}
showing that locomotion results in a MSD that is quadratic in time \cite{Howse2007}. The trajectories of a T-swimmer and its component rods are each plotted for 10s in Fig.\ \ref{fig:emergent_dynamics}d. The MSD shows dramatically enhanced swimming when the rods are assembled into a T-shape (Fig.\ \ref{fig:emergent_dynamics}e). Comparing the MSD for several single rods and T-swimmers (Fig.\ \ref{fig:emergent_dynamics}f), it appears that individual rods have a variety of behaviours: many are diffusive, but some are swimming, which could be due to asymmetries in the rods. The nearly quadratic MSD of T-shaped assemblies indicates a component of deterministic translation, while the change in gradient of in Fig.\ \ref{fig:emergent_dynamics}f shows that the deterministic component $4Dt$ is of a similar magnitude to $v^2t^2$ for this time.

Fitting Eq.\ (\ref{eq:MSAD}) to the MSAD allows the measurement of $\omega$ and $D_r$, while fitting the solution of Eqs.\ (\ref{eq:Langevin_x}-\ref{eq:Langevin_theta}) to the MSD allows the measurement of $v$ and $D$ \cite{Ebbens2010a}. Estimates $D_r$ are in agreement with previous work \cite{Doi1986}. An effective long-time diffusion coefficient can be calculated analytically, as
\begin{equation}
	D_{\text{eff}} = D + \frac{v^2 D_r}{2(D_r^2 + \omega^2)}, \label{eq:d_eff}
\end{equation}
indicating an enhanced diffusivity from motility, with this enhancement lessened by rotation \cite{Ebbens2010a}. Interestingly, measured rotor $D_{\text{eff}}$ are in general smaller than those measured for single rods without fuel and decreases with the number of rods in a rotor. This might be expected as rotors are longer than single rods, which reduces $D$ \cite{Doi1986}. The smaller $D_{\text{eff}}$ does not indicate that $v = 0$, as the enhancement of the translational diffusion is reduced by rotation and negligible if $9v^2 \ll 2L^2(D_r^2 + \omega^2)$ \cite{Doi1986}. 
In fact, rotors move in circular orbits, indicating that $v \neq 0$. The distribution of radii of these orbits is of the same order of magnitude as the distribution of rod lengths, suggesting that the non-zero $v$ is due to variations in the rods.


\section{Self-assembly}

The assembly of rods may be influenced by many physical effects, including hydrodynamic, electric, thermal and steric interactions. For the extensile rods studied here, simulations indicate that hydrodynamics alone is sufficient to bring together pairs of parallel rods \cite{Pandey2014}. To examine the effect of the flow surrounding the rod, experiments were conducted using rods of the alternative configuration, Pt-Au-Pt. These particles are expected to generate a contractile flow, the reverse of Fig.\ \ref{fig:diagram_and_micrographs}a. In experiments with these rods, parallel pairs are not observed to form. This suggests that hydrodynamic attraction is important for bringing the rods into close proximity, revealing the importance of the design of the rods for the self-assembly.

\begin{figure}[tb]
	\includegraphics[width=0.48\textwidth]{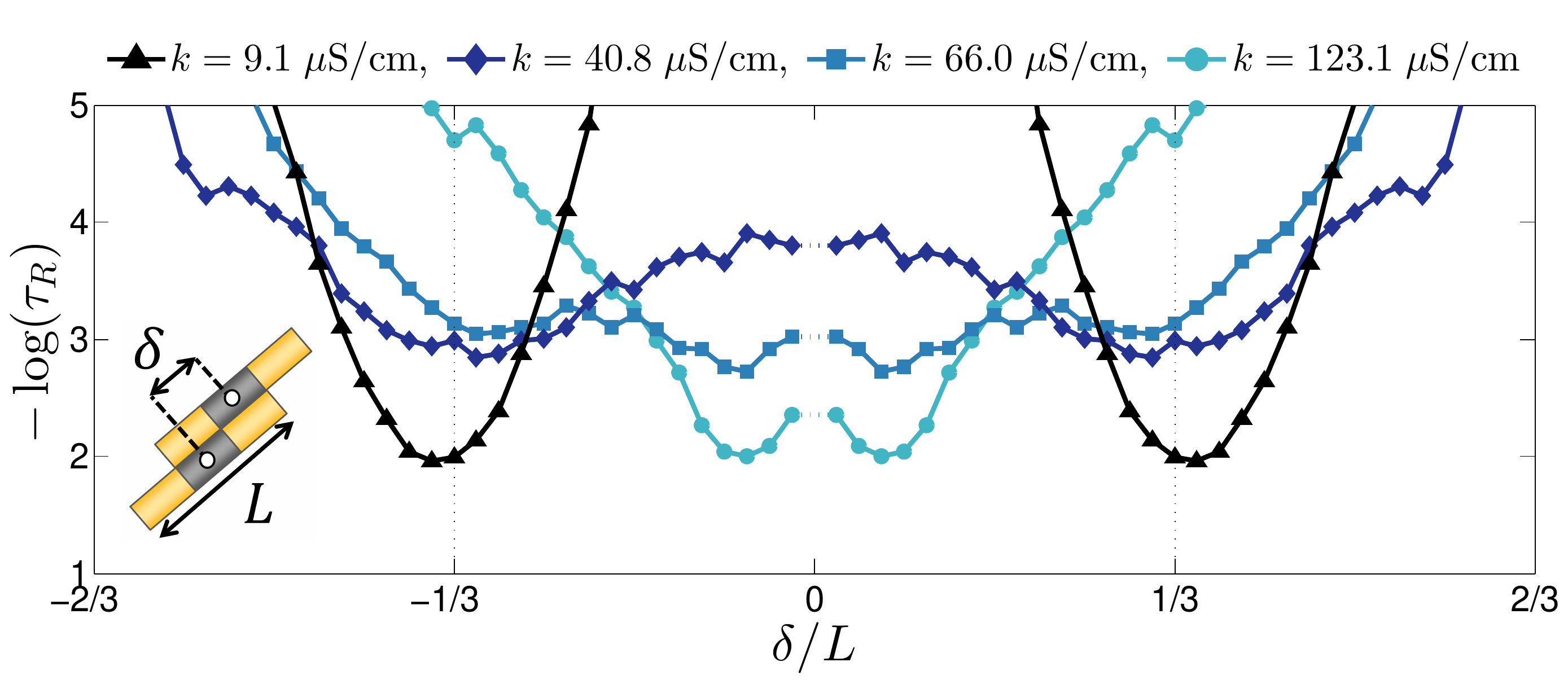}
	
	\caption{\label{fig:energy_diagram} The pseudopotential $-\log(\tau_R)$ as a function of rotor offset $\delta/L$ , where $\tau_R$ is the fractional residence time or probability of an offset. An offset of $\delta \sim L/3$ (see inset) is most prevalent at the lowest conductivity, which corresponds to no added salt. Adding salt increases conductivity, $k$, and leads to a broader distribution of offsets. As only the magnitude of $\delta$ is measured, $-\log(\tau_R)$ is reflected about $\delta = 0$.}
\end{figure}

The detailed configuration of assemblies is influenced by additional effects. In particular, the rotors exhibit a measurable and robust offset between the centers of the constituent rods, as seen in Fig.\ \ref{fig:diagram_and_micrographs}d-f. This observation is quantified by examining the magnitude of the offset $\delta$ between the centroids of paired rods in a rotor (Fig.\ \ref{fig:energy_diagram}, inset) and measuring its variation in time. The fractional residence time $\tau_R$ is the time spent at a particular offset as a proportion of the total time of an experiment. We plot $-\log(\tau_R)$ on Fig.\ \ref{fig:energy_diagram} as, in an analogy with an equilibrium system, the energy of a state $E(\delta) \sim -\log(\tau_R)$. The black curve reveals a minimum centered at $|\delta| = 0.35 L$, corresponding to a preferred overlap of the dissimilar metals (i.e.\ the Pt sections sits alongside Au sections), suggesting that electro-kinetic effects play an important role in determining the configuration. 

These observations suggest that self-assembly is driven by both hydrodynamic and electro-kinetic interactions, which together compete against thermal effects. Changing the conductivity of the solution by the addition of sodium chloride salt (NaCl), weakens both the electro-kinetic and hydrodynamic interactions, thus varying the relative importance of these effects. Specifically, changing the conductivity exerts three main effects: (i) The electric field strength $E$ in the solution has an inverse relationship with conductivity $k$ by Ohm's law, $E = i/k$, where $i$ is the ion flux \cite{Paxton2006}. (ii) The swimming speed of bipartite swimmers has an inverse relationship to solution conductivity $U \sim 1/k$, where $U$ is the swimming speed \cite{Moran2014} and a similar relationship for the flow around the tripartite rods would be expected. (iii) The length scale $\lambda$ of the near-surface region where flow and electric field occur, scales as $\lambda \sim 1/k$.

Due to the weakened interactions, it is expected that the offset between rods will be more widely distributed. Experiments were conducted at $5\%$ H$_2$O$_2$ and four values of the conductivity (corresponding to NaCl concentrations of $0$, $0.17$, $0.44$, and $0.87$ mM), again measuring the variation in the magnitude of $\delta$, here by examining the change in perimeter of a rotor over time (SI). The results are plotted as pseudopotentials in Fig.\ \ref{fig:energy_diagram}, where the black curve corresponds to no added salt. Increasing the conductivity results in the component rods sliding parallel to each other, causing large variations in $\delta$ over time and thus a more uniformly distributed pseudopotential.


\section{Discussion}

The formation of both rotors and T-swimmers from symmetric active particles have been reported in a computational simulations of filaments constructed of active beads producing force-free, torque-free flows in a fluid, confined to move in 2D in an infinite domain \cite{Pandey2014}. Filaments surrounded by an overall extensile flow (similar to what is expected for the rods studied here) formed into parallel pairs which rotate, but in the opposite direction to that which is observed in our experiments (i.e.\ in these simulations R1 would rotate clockwise, R2 counter-clockwise). The formation of T-shaped swimmers was also predicted by these simulations, but for particles surrounded by contractile flows \cite{Pandey2014}. Notably, he T-swimmers in our experiments were formed from particles inducing extensile flows in their surroundings and ordered assemblies were not observed to form in experiments using particles with contractile flows.

The mechanism for motion of assemblies could be explained if the flow on one side of a rod in a rotor is reduced by the presence of the other rod, leaving only the outer flows. In this case, the flow induced by the assembly results in a net rotation in the fluid due to the offset between the rods. By conservation of angular momentum, the rotor must rotate in the opposite direction. This mechanism correctly predicts the rotation direction. Similarly, if the presence of the rod that makes the leg (vertical rod in Fig.\ \ref{fig:diagram_and_micrographs}c) in a T-swimmer reduces the flow on one side of the rod that makes the top, then the net result of the flows will be to pump fluid in a direction aligned with the leg of T-shaped assembly, causing it to swim towards the junction end.

We have described the dynamic and reversible self-assembly of rotors and T-swimmers from symmetric, tripartite Au-Pt-Au micron-scale rods. Self-assembly breaks the symmetry of individual particles, leading to directed motion. This system has rich dynamics, resulting from the complex interplay of various forces, with the potential for control over the rotation rate, swimming velocity and lifetime of assembled structures through the fuel concentration, solution conductivity and tailoring of particle geometry and composition. This is a rare example of artificial dynamic self-assembly, a process common in nature. Systems of this type are of interest not only because of potential applications in micro-fluidics but also because a more detailed understanding of dynamic self-assembly could lead to insights about similar processes in nature.

\section{Acknowledgements}

We thank Y.\ Liu, A.\ Hollingsworth and M.\ Driscoll for useful conversations. This work was primarily supported by the Materials Research Science and Engineering Center (MRSEC) program of the National Science Foundation under Award Number DMR-1420073. M.S.D.W.\ thanks the Fulbright Scholarship Lloyds of London Award for financial support. T.A.\ thanks the JSPS Postdoctoral Fellowships for Research Abroad for the financial support.

\bibliography{active_matter}

\end{document}